\documentclass[pre,a4paper,superscriptaddress,nofootinbib,twocolumn]{revtex4}

\usepackage{natbib}
\usepackage{epsfig}
\usepackage{graphicx}

\def\x{{\mbox{\boldmath$x$}}}
\def\y{{\mbox{\boldmath$y$}}}

\def\bomega{{\mbox{\boldmath$\omega$}}}

\def\nab{{\bf \nabla}}

\def\begineq{\begin{equation}} 
\def\endeq{\end{equation}}
\def\be{\begin{equation}}
\def\ee{\end{equation}}

\begin{document} 

\title[The evolution of energy in flow driven by rising bubbles]
{The evolution of energy in flow driven by rising bubbles}
\author{Irene M. Mazzitelli and Detlef Lohse}
\affiliation{Physics of Fluids Group, \\
Department of Science and Technology, J.\ M.\ 
Burgers Centre for
Fluid Dynamics, and Impact Institute\\ University of Twente, P.O. Box 217, 7500 AE Enschede, The Netherlands}
\date{\today}


\begin{abstract}
We investigate by direct numerical simulations
the flow that rising bubbles cause in an
originally quiescent fluid.
We employ the Eulerian-Lagrangian method with 
two-way coupling and periodic boundary conditions. 
In order to be able to treat up to 288000 bubbles,
the following  approximations and simplifications
had to be introduced, as done before e.g.\ 
by Climent and Magnaudet, Phys.\ Rev.\ Lett.\ 82, 4827 (1999):
(i) The bubbles were treated as point-particles, thus (ii) disregarding
the near-field interactions among them, and (iii) 
effective force models for the lift and the drag forces were used. 
In particular, the lift coefficient was assumed to be 1/2, independent
of the bubble Reynolds number and the local flow field.
The results suggest that large scale motions 
are generated, owing to an {\it inverse
energy cascade} from the small to the large scales. However, as the Taylor-Reynolds number
is only in the range of 1, the corresponding scaling of the energy spectrum with an exponent of $-5/3$ cannot develop over a 
pronounced range.
In the long term, the
property of local energy transfer, characteristic of
real turbulence, is lost and the input of energy equals the 
viscous dissipation at all scales.
Due to the lack of strong vortices the bubbles spread rather
uniformly in the flow. The mechanism for uniform spreading is as 
follows: Rising bubbles induce a velocity field behind them that
acts on the following bubbles. Owing to the shear, those bubbles 
experience a lift force which make them spread to the left or right,
thus preventing the formation of vertical bubble clusters and
therefore of efficient forcing. Indeed, when the lift is artifically put to
zero in the simulations, the flow is forced much more efficiently
and a more pronounced energy accumulates at large scales (due to the 
 inverse energy cascade) is achieved. 
\end{abstract}

\maketitle

\section{Introduction}

Bubbly flows are ubiquitous in nature and technology,
but exact analytical or numerical results are extremely difficult to obtain,
due to the multiscale nature of the problem and due to the
moving interfaces. For a review on the numerical modeling we refer
to 
\cite{cro96,mag00}, for recent reviews on the experimental situation
we refer to  \cite{mud05,mar07}, and our own recent work on the subject
has been summarized  in \cite{ber06b}.
An excellent overview on the experimental, numerical, and theoretical
knowledge for various bubble Reynolds numbers can also be found in refs.\
\cite{bun02a,bun02b}.

The motion of the small bubbles in the fluid induces 
velocity fluctuations that can be either dissipated immediately
by viscosity or can be enhanced, thus generating
motion on scales much larger than the disturbance dimension.
Owing to their random character, these fluctuations are refered to as 
``pseudo-turbulence'' \cite{wij98,bun02b,mar07}.
In a
flow initially at rest and only forced by rising bubbles,
the pseudo-turbulent
fluctuations are the only source of energy. 
Otherwise they can add to the
already existing fluid velocity fluctuations, which are driven in
some other way. 
Depending on the flow conditions and 
the bubble size distribution, the turbulent energy dissipation may be either 
enhanced or suppressed. 

Lance and Bataille \cite{lan91} have suggested that the effect
of bubbles on the flow depends on the ratio of kinetic energy due
to the bubble motion and the typical kinetic energy
$\left<u'^2\right>$ of the fluctuations in the liquid velocity before bubble 
injection,
\begin{equation}
    b =  \frac{1}{2} \alpha v_T^2  \, / \,  \left< u'^2 \right >,
\end{equation} where $u'$ is the typical flow velocity fluctuation, $\alpha$
the void fraction, $v_T$ the bubble rise velocity in still water, 
and we have taken ${1}\over{2}$ for the added mass coefficient.  The ratio 
 $b$ is called the bubblance parameter \cite{ber06b}. 
For $b \ll 1$ the kinetics of the
bubbly flow are entirely dominated by the turbulent energy of the
flow and the bubbles can be considered as some distortion, such
as in the experiments of refs.\ 
\cite{the82,tsu82,tsu84,mud97,ren05,ber06c} or in the numerical
simulations of refs.\ \cite{max96b,dru98,dru01a,xu02,maz03a,maz03b},
whereas for $b \gg 1$ the flow is dominated by the bubble motion; i.e.\ we
have bubblance rather than turbulence, such as in the experiments of ref.\
\cite{lan91,zen01,car01,mud01,ris02,mar07} or in the numerical simulations of ref.\ \cite{esm98,esm99,cli99,bun02a,bun02b,esm05}. 
The analogous situation for sedimenting particles has experimentally been analyzed by Faeth and coworkers, both for
particles in water  \cite{faeth} and in air \cite{mizu}.


In the present work we focus on the bubblance  case $b\gg 1$, namely on
microbubbles rising in an initially
quiescent flow, where formally $b=\infty$. 
These conditions imply that pseudo-turbulence, due to the
bubbles' buoyancy, is the only source of flow energy, thus bubbles
drive the turbulence and eventually the energy dissipation. 
We will address the following questions: (i) 
What is the time evolution of the energy of bubbly driven 
turbulence initially at rest?
(ii) Are microbubbles able to 
induce in still fluid a flow that possesses similar features 
as real turbulence, i.e., can the inertial scaling law characteristic
of homogeneous and isotropic turbulence 
be attained in a flow forced solely by bubbles?
(iii) How are bubbles eventually distributed in such flow and what forces 
determine this distribution?

To be able to address these questions, the flow should best 
be driven at least
by ten-thousands of bubbles, in order to have sufficient statistics.
 While hundreds of bubbles can still be treated
with front-tracking techniques \cite{esm98,bun99,esm99,bun02a,bun02b,esm05}, 
resolving both the shape of the bubble and the flow around it,
this clearly is no longer possible for ten-thousands of bubbles, and one therefore
is forced to employ approximations. We will thus follow the complementary
approach 
and model the bubbles as  {\it point-particles}, 
on which effective forces such as drag and lift act \cite{mag00},
similarily as has also been done by 
Climent and Magnaudet \cite{cli99}.
We thus disregard 
the near-field interactions among 
the bubbles. 
Also the effective forces had to be modelled, namely by chosing 
drag and lift coefficients. Unfortunately, the lift coefficient is not
well known \cite{mag00}. In fact, it depends on the bubble Reynolds number
and on the local shear and vorticity in a highly non-trivial way 
\cite{sri95,leg98,tom02,mer05,cli06,nie07}. 
Moreover, due to the interactions with the wake,
effective forces on bubbles can even be non-local in time (history forces)
\cite{mei94,leg98,mag98}. Given these complications, for conceptional
simplicity we decided to simply use the Auton  lift coefficient  $C_L = 1/2$
\cite{aut87}, 
realizing that at best {\it qualitative} agreement between our simulations
and possible experiments can be expected. 
Finally, periodic boundary conditions are employed. 
The aim of  the present paper can thus 
only be
to identify mechanisms; no quantitative predictions or comparisons with 
experiments are possible. 
The numerical simulations for bubble-induced convection
 \cite{cli99} or  for Taylor-Couette flow with microbubbles 
inducing drag reduction \cite{sug08b} were done in the same spirit with
related numerical schemes, and indeed the relevant physical mechanisms could
be identified.

Our numerical simulations  are
 based on the code extensively described in \cite{maz03b}
but now as stated above 
with forcing only by the bubbles, to mimick the 
pseudoturbulent flow. For completeness we
briefly repeat the dynamical equations and the 
central assumption in section \ref{sec2}.
Sections \ref{sec3} and  \ref{sec4} describe
the time evolution  of global and spectral observables, respectively.
In section \ref{sec5} we propose a qualitative 
physical explanation for the 
detected fluid energy time evolution.
Section \ref{sec6} contains conclusions.

{\section{Bubbles and fluid equations}\label{sec2}}
\subsection{Bubble motion equation}
The motion of a small, undeformable, 
bubble embedded in a velocity field ${\bf u} (\x ,t)$ can 
be modelled by the equation (see e.g.\ \cite{tho84,cli99,mag00}):
\begin{eqnarray}
\label{eq1}
{d {\bf v} \over dt} = 3 { D {\bf u} \over Dt} + {1 \over \tau_b}
[ {\bf u} (\y(t),t) - {\bf v}(t)] - 2 {\bf g} \\ \nonumber
- [{\bf v}(t)-{\bf u}(\y(t),t)] \times
\bomega (\y(t),t).
\end{eqnarray}
The various symbols denote: $\y(t)$ the bubble position, ${\bf v}$
the bubble velocity, $\bomega  = \nab \times {\bf u}$
the fluid vorticity,
${\bf g}$ the gravity (acting in negative z-direction), and $\tau_b$
the relaxation time, i.e.\ the time needed to 
adjust the bubble velocity to that of the fluid.
The latter is related with the terminal velocity $v_T$  
in still fluid, $v_T=2g \tau_b$.
In the case of small bubble Reynolds number
$Re = 2|{\bf u} - {\bf v}| a/\nu < 1$, 
with $a$ the bubble radius and
$\nu$ the kinematic viscosity, for the bubble 
response time it holds $\tau_b = a^2/6\nu$
\cite{had11,ryb11}. For larger bubbles the prefactor in this relation
is larger than 1/6, which however would only quantitatively affect
our results.  
The material derivative ($D/Dt$) of the fluid
velocity is evaluated at the bubble position.
Eq.\ (\ref{eq1}) embodies the effect of
fluid acceleration plus added mass, drag force,
buoyancy, and lift force, with the lift coefficient set to
 $C_L = 1/2$ for simplicity, as discussed in the introduction.
We refer to refs.\ \cite{max83,mag00,maz03b} for a 
detailed description of the terms in eq.\ (\ref{eq1}).

{\subsection{Simulation of the flow}}
The computational domain is a cube of 
side $L_0=2\pi$, consisting of $N^3 = 128^3$ mesh points 
and subjected to periodic boundary conditions.
The simulation is started at $t=0$ with the flow 
at rest and with  $N_b$ 
bubbles with $Re \sim O(1)$ placed at random
locations. Bubbles rise because of gravity 
and transfer momentum to the fluid. 
We track their trajectories and
we treat each bubble as a point-source of momentum. 
Then, the total action on the flow results 
by summing the 
$\delta$-forcing that the bubbles apply
at their positions  \cite{cli99,boi98a,maz03b}:
\begin{equation}
{\bf f}_b (\x,t) = \sum_{i=1}^{N_b} 
\Big({D \over Dt} {\bf u} (\y_i(t),t) - {\bf g}\Big) 
\left({4\pi \over 3}a^3 \right) \delta (\x - \y_i(t)).
\label{bubbleforcing}
\end{equation}
Here ${\bf g}$ is the gravity, directed along the negative $z$ axis.
The induced flow velocity ${\bf u}(\x,t)$ evolves according 
to the incompressible Navier-Stokes equation:
\begin{equation}
\label{NSeqs}
{\partial {\bf u} \over \partial t} + {\bf u}\cdot \nab {\bf u}
= - {\bf \nabla} p + \nu \Delta {\bf u} + {\bf f}_b
\end{equation}
which is solved by direct numerical simulation.

\begin{table}
\begin{center}
\begin{tabular}{|c|c|c|c|c|c|c|c|}
\hline
 & $\alpha$ & $N_b$ & Re & $\tau_b$ & $a~~(\mu m)$&$E_p\cdot 10^3$&$E\cdot 10^3$\\ 
\hline
(a) & $1.6 \%$ & $144000$ & 3.10 & $8.4 \cdot 10^{-3}$  & $78$ & $1.34$ &  $1.96$\\
(b) & $0.8 \%$ & $72000$  & 3.10 & $8.4 \cdot 10^{-3}$  & $78$ & $0.67$ &  $1.38$\\
(c) & $3.2 \%$ & $288000$ & 3.10 & $8.4 \cdot 10^{-3}$  & $78$ & $2.67$ &  $2.25$\\
(d) & $1.6 \%$ & $50848$  & 4.37 &$16.7 \cdot 10^{-3}$ & $87$ 
& $1.34$ &  $1.57$ \\
\hline
\end{tabular}
\end{center}
\caption{Simulation parameters for all cases 
analysed: void fraction $\alpha$,
total bubbles number $N_b$, bubble Reynolds number $Re = 2v_T a/\nu$,  
bubble response time $\tau_b$,  
equivalent radius $a$ in physical units, potential flow estimate of the
total energy induced in the flow $E_p= \alpha v_T^2/4$, and time
asymptotic estimate from our calculations 
of the total energy $E$. This value has been determined from time-averaging the 
statistically stationary state.
In the numerical simulations
 the kinematic viscosity $\nu=0.007$, $\alpha$,
the rise speed in still fluid $v_T=0.578$, and $\tau_b$ 
are fixed, the other quantities result consequently. In particular, 
the intensity of the
gravity results from $g=v_T/2\tau_b$ and therefore is different in the 
simulation $(d)$ as compared to  the other ones.} 
\label{table1}
\end{table}

\begin{table}
\begin{center}
\begin{tabular}{|c|c|c|}
\hline
 \hskip .5truecm & {\it dimensionless  } & {\it physical  } \\
 \hskip .5truecm & {\it parameter  }     & {\it equivalent  } \\ 
\hline
$~~~\nu~~~$ & $0.007$ & $10^{-2}cm^2/s~~$ \\
$~~~g~~~$   & $34.55$ & $981cm/s^2~~$ \\
$~~~a~~~$   & $0.019$ & $78 \mu m~~$ \\
$~~~v_T~~~$ & $0.58 $ & $2cm/s~~$ \\
$~~~\tau_b~~~$ & $8.4 \cdot 10^{-3}$ & $1ms~~$ \\
\hline
\end{tabular}
\end{center}

\caption{Simulation parameters for cases $(a)$ to $(c)$ and
corresponding physical equivalents. The bubble Reynolds number
is $Re= 2v_T a /\nu = 3.10$. 
}

\label{table2}
\end{table}

We use the pseudo-spectral code
described in \cite{maz03b} where also
the other details on the numerical simulations can be found. 
The point force approximation 
is validated by performing the same tests as in 
\cite{maz03b}. 
We stress that, in contrast to that
earlier work of ours, here eq.\ (\ref{NSeqs}) does 
not contain any forcing
on the large scales: the flow is sustained solely by the  
bubble forcing term ${\bf f}_b$.

We analyse different cases. A list of the flow and bubbles
parameters is shown in Table \ref{table1}. 
In Table \ref{table2} the values of the 
numerical parameters and their physical 
equivalents are presented for some of the simulations performed.

\section {Evolution in time for global quantities}\label{sec3}
We describe the energy time evolution of the pseudo-turbulent 
field generated by  
the rise of microbubbles. 
As already stated, at time $t=0$, the bubbles are 
randomly placed in the flow, that is originally at rest.
Their rise displaces liquid and thus generates velocity fluctuations. 
If these fluctuations are not 
rapidly dissipated by viscosity,
they can be transmitted 
to larger scales. As a consequence, large scale motions 
are produced and the flow may become turbulent.

We investigate this issue by measuring 
average flow quantities as well as by studying
the spectral energy distribution. We focus on case $(a)$ 
of Table \ref{table1}.

In Fig.\ \ref{fig1}$a$ we plot the total fluid energy, 
$E(t)=\langle u_x(t)^2 +u_y(t)^2 +u_z(t)^2 \rangle/2$,
as a function of time.
In the beginning $E(t)$ 
undergoes a steep rise, afterwards it slowly decreases,
until it begins to oscillate and a
statistically stationary state is reached.
The behavior qualitatively resembles that one of the front-tracking simulations
of ref.\ \cite{bun02b}, see figure 1 of that paper.  
The kinetic energy is mainly generated by the momentum transfer 
in the direction of gravity, as we confirm by measuring the three 
components of the fluid velocity fluctuations
$ \langle u_i^2 \rangle$, with $i=x,y,z,$ 
which are much larger in the $z$ direction than in the
horizontal ones,
and the unidimensional
Taylor-Reynolds number, defined as:
$$Re^i_\lambda = \sqrt{ \langle u_i^2 \rangle \over 
\langle \partial_i u_i^2 \rangle}{\sqrt {\langle u_i^2 \rangle} \over \nu}
\qquad \hbox{(no sum over i)}.$$  
The behavior of $Re_\lambda^i$ as function of time is presented 
in Fig. \ref{reynolds_eps}. It is evident from the plot that 
the flow displays strong asymmetry.
 
We now compare these results with potential flow theory, see e.g.\ 
the work of van Wijngaarden \cite{wij82,wij98,wij05}. We do not expect
agreement, given that we deal with point-particles and that
the potential flow  results of refs.\ 
\cite{wij82,wij98,wij05}
only hold in the 
 high bubble Reynolds number case, and here
we have $Re$ between 3 and 5. 
Nonetheless, 
we consider this comparison to be instructive and
surprisingly we find 
the saturated kinetic energy to be of order $\alpha v_T^2/4$, 
in agreement with 
the potential flow theory results 
for high $Re$ bubbles.
However, the 
redistribution of the energy along
the three directions deviates from what is
predicted by potential flow analysis,
according to which we should have \cite{wij82}:
$$\langle u_z^2 \rangle \simeq {1 \over  5}\left( 
\alpha {v_T^2 }\right),~~~ \langle u_x^2 \rangle = 
\langle u_y^2 \rangle \simeq {3 \over 20} \left( 
\alpha {v_T^2 }\right)$$ 
The potential flow value is  $\langle u_z^2 \rangle / 
\langle u_x^2\rangle \simeq 4/3$, whereas in our simulation this 
ratio is about $15$.
In the opposite limit, $Re \to 0$,
under Stokes flow condition, the fluid equations are
linear. In this regime, symmetry
considerations require that rising bubbles
cannot force the flow in directions perpendicular to their motion
(provided that there are no walls but periodic boundary conditions
as in our case),
a result that is also intuitive for rising point particles 
in fluid at rest.
As a consequence the ratio  $\langle u_z^2 \rangle / 
\langle u_x^2\rangle \to \infty$. Our result, for small but
finite Reynolds number,
lies in between the two limits discussed.
The same holds for the case of sedimenting particles in water (with larger particle Reynolds numbers 
between 38 and 545), where
this ratio is between 4 and 25 \cite{faeth}.

The total energy induced by high $Re$ bubbles,
for which inertia effects are dominant, can be easily
estimated by the following argument: 
at low void fractions the flow energy is the sum of the
energy induced by individual bubbles, i.e., $E\simeq N_b (1/2)m_bv_T^2$,
where the effective {\it mass} of a bubble is $m_b = \rho_f (2 \pi a^3/3)$,
owing to the added mass factor $1/2$. Thus $E\simeq \alpha v_T^2/4$.
On the other hand, when $Re \sim 1$, 
to estimate the total energy induced by
sedimenting particles or rising bubbles is
far more complicated. Indeed, for $Re \ll 1$, the flow induced by
one particle decreases, with the distance $r$ from 
the particle, as $1/r$, 
thus leading to a total flow energy
that diverges with the system size. Different screening mechanisms 
have been invoked in the past in order to account for this problem
(see e.g.\  \cite{koc91,koc93,nic95a,nic95b,bre99,ngu05}), but this issue
is beyond the scope of the present paper.

In Table \ref{table1} we report the total energy estimated in 
our simulations, 
correspondent to different void fractions and bubble dimensions.
In all cases the energy induced is of order $\alpha v_T^2/4$.
We note, by comparing cases $(a)$ and $(d)$,
that, when increasing the bubble dimension while fixing the void 
fraction, thus reducing the total bubble-fluid interface,
less energy is generated in the flow.
\section{Evolution in time for spectra}\label{sec4}
After transforming to wavenumber space we consider
the time development of the energy spectrum:
\begin{equation}
E(k,t) = {1 \over 2} \sum_{k<|{\bf k}| < k+dk} 
u^*_i({\bf k},t)u_i({\bf k},t)~~~i=x,y,z.
\label{spectrum}
\end{equation}
Here $u_i({\bf k},t)$ is the $i$th component 
of the fluid velocity in $k$-space and 
repeated indices are considered summed.
$E(k,t)$ is the total energy contained in a spherical 
shell of radius $k$ and width $dk$.

\begin{figure}
\begin{center}
\includegraphics*[width=5cm]{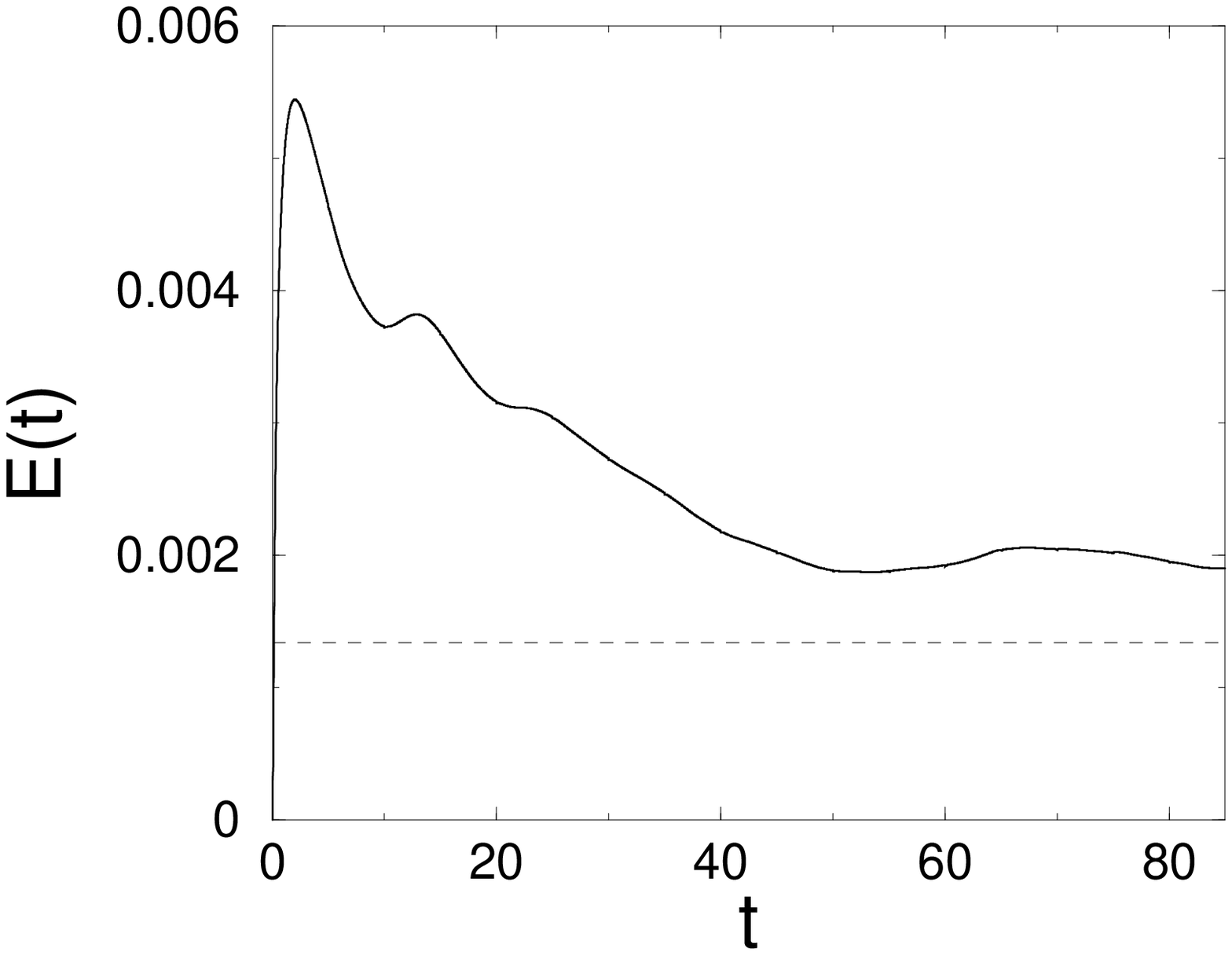}
\includegraphics*[width=5cm]{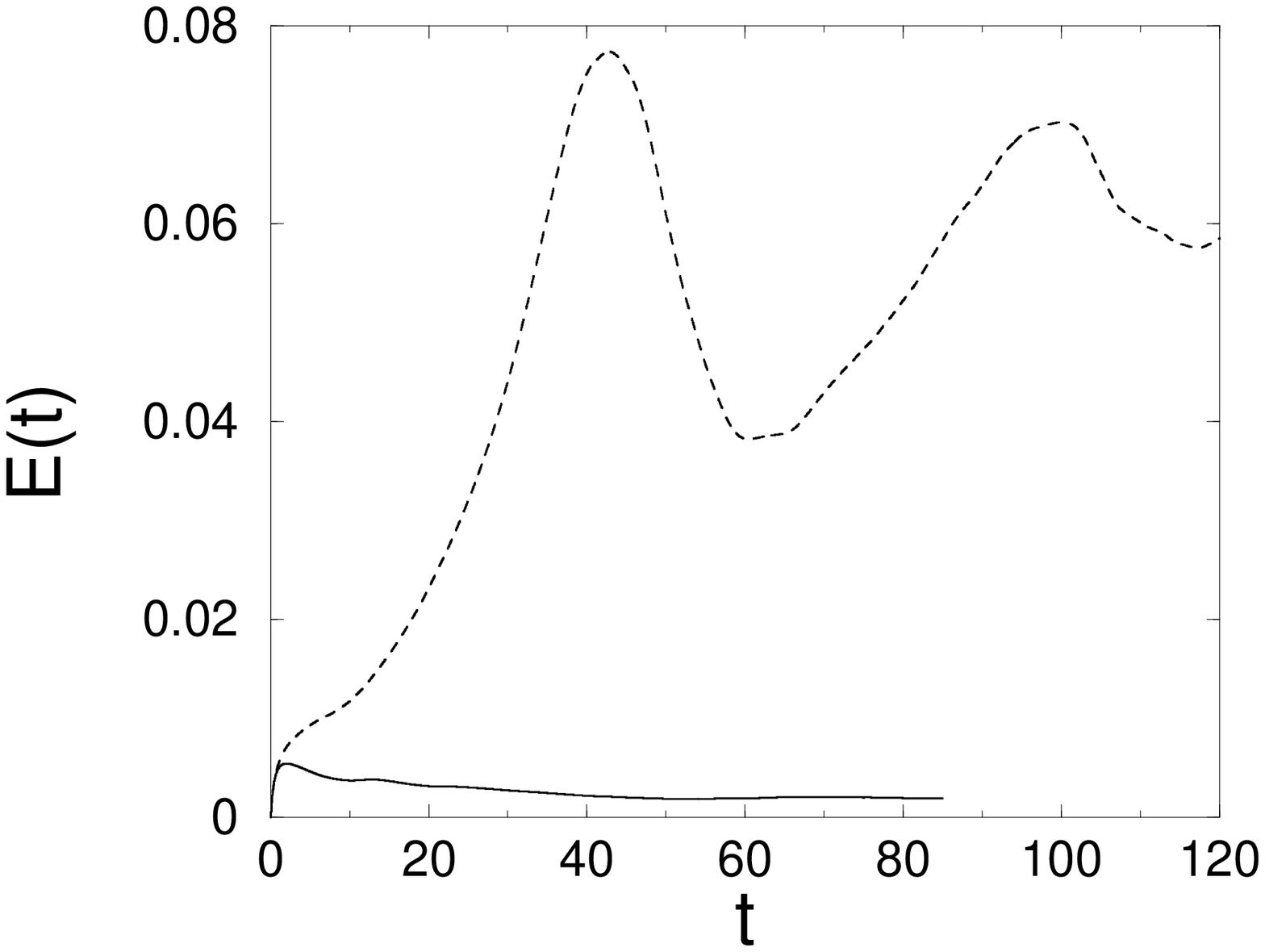}
\caption{$(a)$  Total fluid energy as function of time. The straight dashed line indicates the potential flow result: $E = \alpha v_T^2/4$.
$(b)$  Total fluid energy in simulation with lift (solid line) and 
without (dashed line) as functions of time.}

\label{fig1}
\end{center}
\end{figure}

\begin{figure}
\includegraphics*[width=5cm]{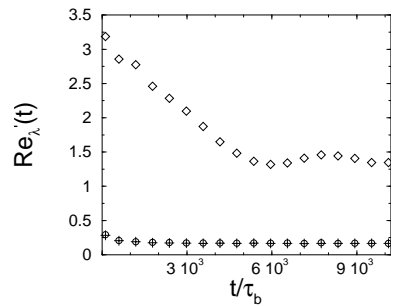}
\caption{Behavior of the uni-dimensional Taylor-Reynolds number as function of time, in the $x$-(pluses), $y$-(circles), and $z$-(diamonds) directions.}
\label{reynolds_eps}
\end{figure}

\begin{figure}
\begin{center}
\includegraphics*[width=5cm]{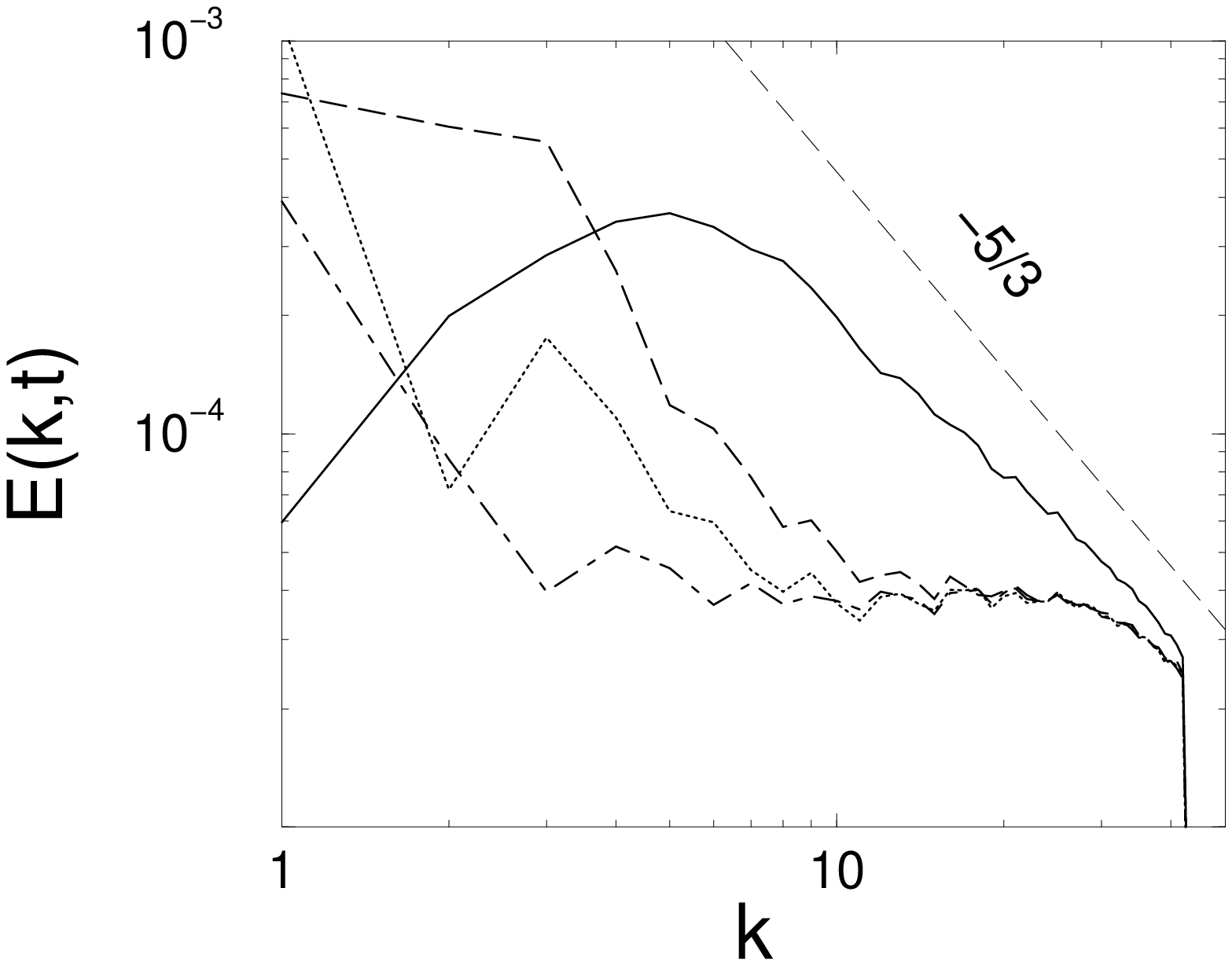}
\vspace*{0.25cm}
\includegraphics*[width=5cm]{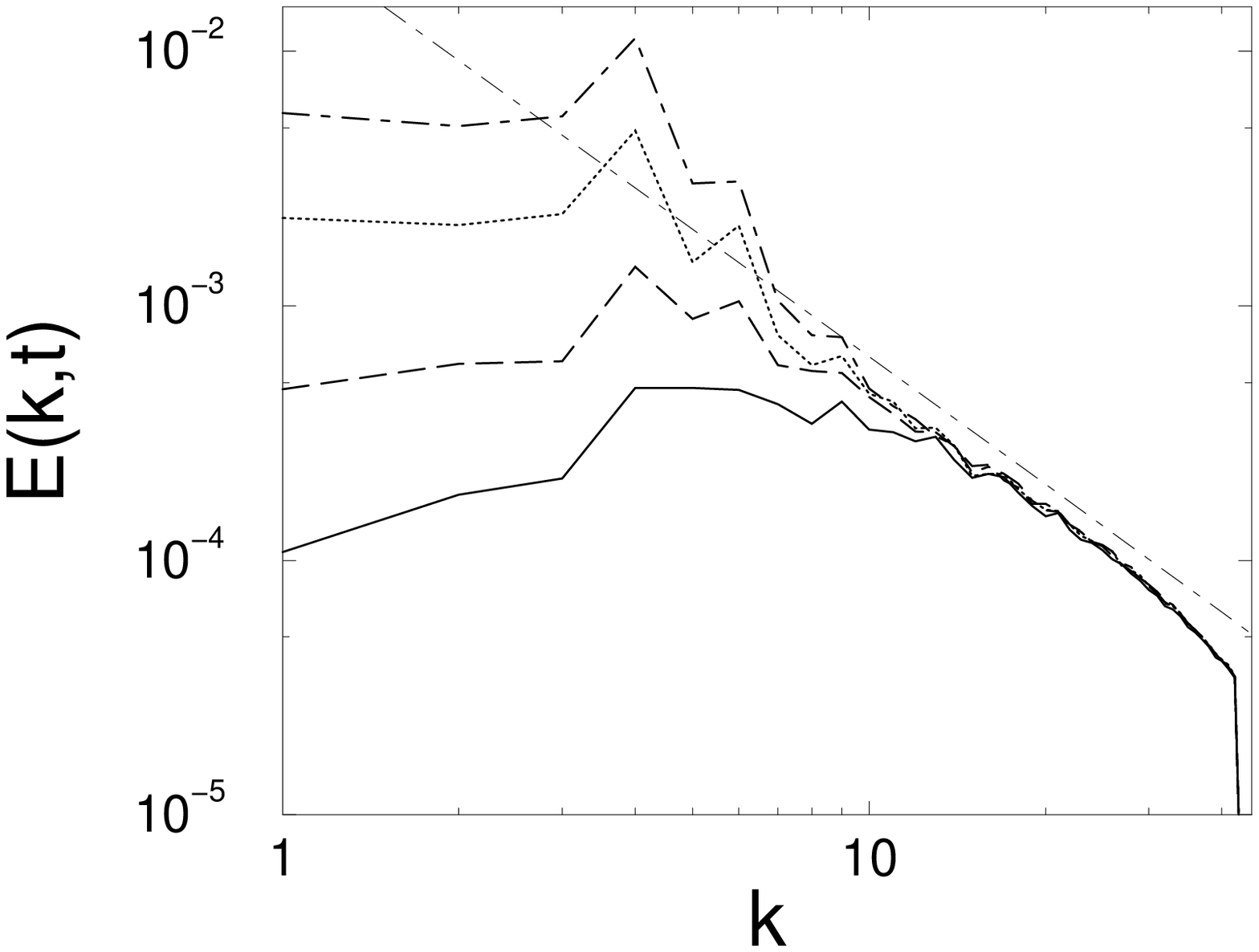}
\caption{$(a)$  Energy spectra for the simulation that includes lift
forces obtained by averaging on four different 
time intervals: $0<t/\tau_b<6 \cdot 10^2$ (solid line), $12 \cdot 10^2<t/\tau_b<18 \cdot 10^2$ (dashed line), $24 \cdot 10^2<t/\tau_b<30 \cdot 10^2$ 
(dotted line), and $100 \cdot 10^2<t/\tau_b<106 \cdot 10^2$ (dot-dashed line). 
The straight line indicates the behavior in homogeneous and isotropic turbulence. $(b)$  Energy spectra for the simulation without lift forces obtained by averaging on four different 
time intervals. For the various symbols look at the caption of Fig. \ref{fig3}$a$.}
\label{fig3}
\end{center}
\end{figure}


In Fig.\ \ref{fig3}$a$ the energy spectrum, 
averaged on four subsequent time 
intervals, is presented. The intervals correspond to  
$0<t/\tau_b<6 \cdot 10^2$, $12 \cdot 10^2<t/\tau_b<18 \cdot 10^2$, $24 \cdot 10^2<t/\tau_b<30 \cdot 10^2$,
and $100 \cdot 10^2<t/\tau_b<106 \cdot 10^2$
in Fig. \ref{fig1}. 
The figure depicts that the energy is originally 
introduced at high wavenumbers and gradually 
transported to larger scales (solid line). 
However, after some time the spectrum flattens and 
a nearly constant energy is measured at all scales.
Thus, in the first stage of bubble-fluid coupling, 
an {\it inverse energy cascade}, from the 
small to the large scales, builds up large scale eddies. The corresponding slope of the
spectrum -- indicating the local transfer of energy -- would be  $-5/3$, however, as the Taylor-Reynolds numbers are small,
such a scaling regime cannot really develop.
Later on, the inverse energy cascade 
cannot be sustained and it finally disappears in the final state.
We investigate whether this last state is statistically stationary by having
a closer look at the energy transfer equation in $k$-space:
\begin{equation}
{\partial \over \partial t} E(k,t) = T(k,t) - D(k,t) + F_b(k,t)
\label{energytransfer}
\end{equation}
Here, the various terms indicate, respectively:
the energy transfer to
wavenumber $k$, $T(k,t)$,
$$T(k,t) = \sum_{k < |{\bf k}| < k+dk} T({\bf k},t),$$
where
\begin{equation}
T({\bf k},t) = {\cal I}m\left(k_j u^\ast_l({\bf k},t)\sum_{\bf k'} u_j({\bf k}
- {\bf k'},t) u_l({\bf k'},t)\right),
\end{equation}
the viscous dissipation, $D(k,t) = 2 \nu k^2 E(k,t)$, and
the bubble forcing contribution, $F_b(k,t)$,
$$F_b(k,t) = \sum_{k < |{\bf k}| < k+dk} F_b({\bf k},t),$$
where
\begin{equation}
F_b({\bf k},t) = {\cal R}e \Big( {\bf u}^*({\bf k},t) \cdot {\bf {\tilde f}}_b({\bf k},t) \Big),
\label{spe_forc}
\end{equation}
with ${\bf {\tilde f}}_b({\bf k},t)$ the Fourier 
transform of the coupling term
${\bf f}_b (\x,t)$, defined in eq.\ (\ref{bubbleforcing}). 
Here ${\cal I}m$ and 
${\cal R}e$ indicate, respectively, the imaginary and real 
part of the expression between the brackets.

\begin{figure}
\begin{center}
\includegraphics*[width=5cm]{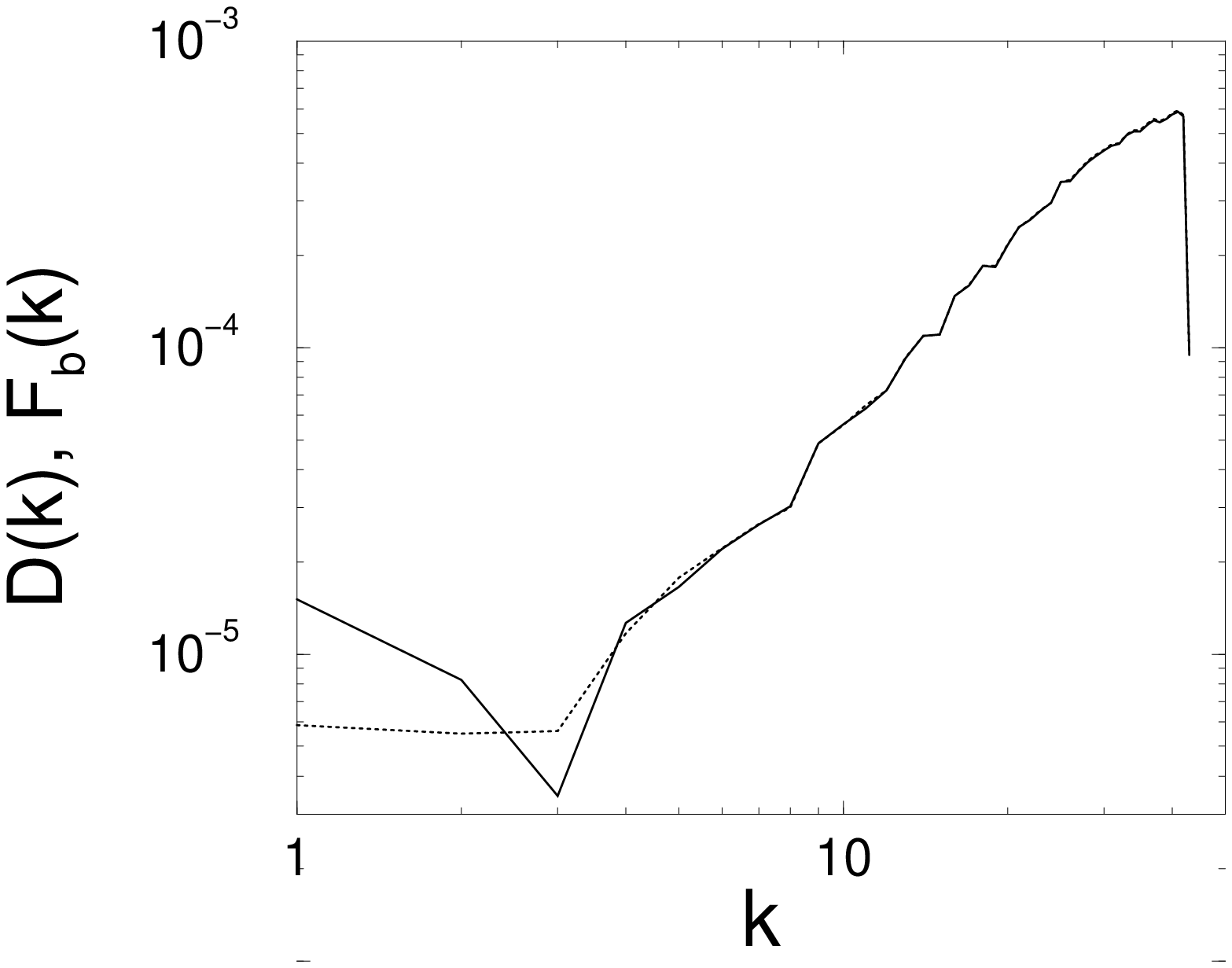}
\includegraphics*[width=5cm]{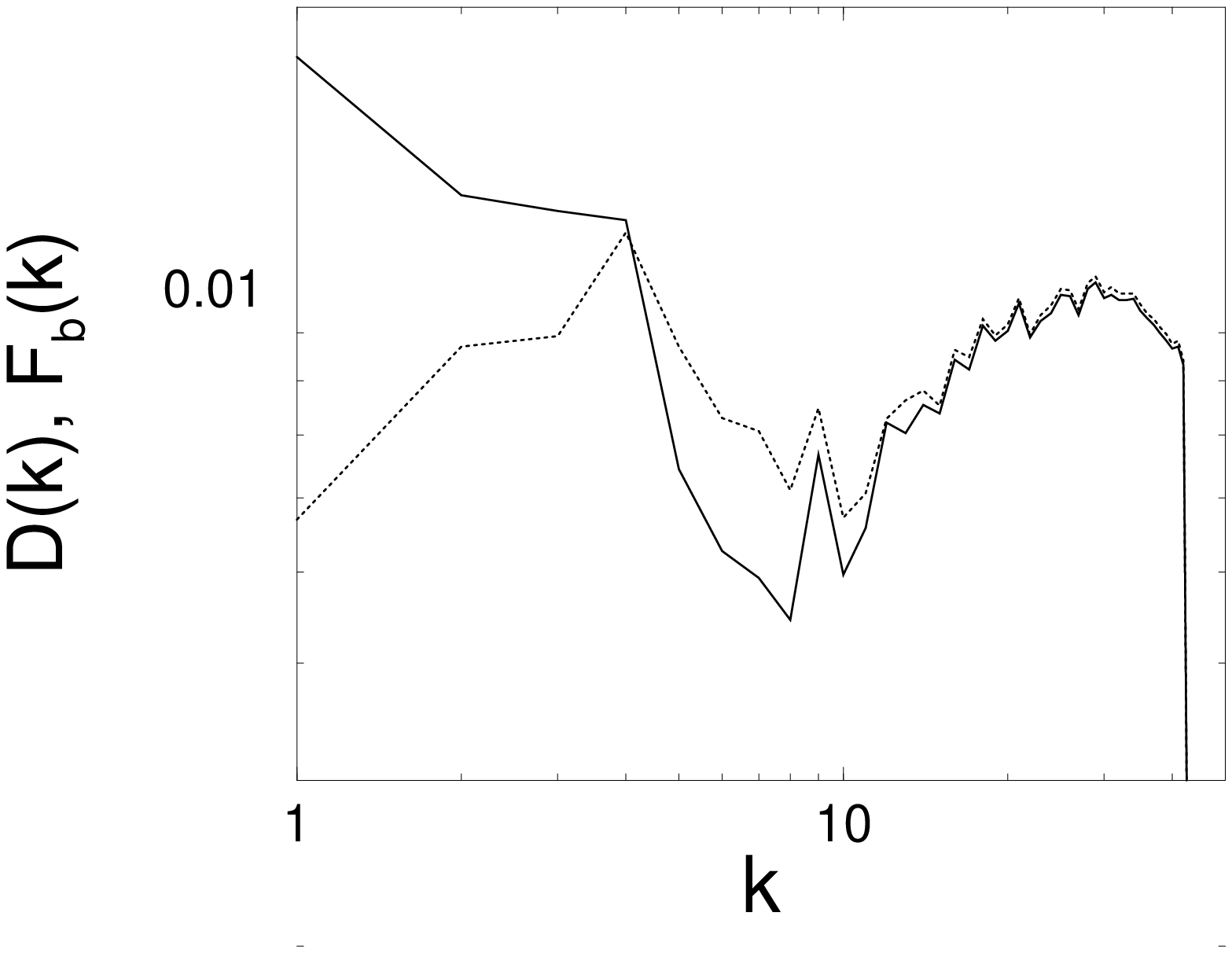}
\caption{$(a)$  Time average of the contribution of the bubble forcing to the energy spectrum, as defined in eq.~(\ref{spe_forc})
 (solid line) and of the viscous energy 
dissipation $D(k)=2\nu k^2 E(k)$ (dotted line), in the simulation 
with lift force. $(b)$  Time average of the contribution of the bubble forcing to the energy spectrum 
(solid line) and of the viscous energy 
dissipation $D(k)=2\nu k^2 E(k)$ (dotted line), in the simulation 
without lift force.}
\label{spe_forc_eps}
\end{center}
\end{figure}

The spectra of the bubble forcing and of the dissipation are 
shown in Fig.~\ref{spe_forc_eps}$a$. The strongest forcing is concentrated on 
the small scales, as we expect, owing to the dimension of the 
energy sources. However, the energy that is 
initially transfered to the 
large scales via the nonlinear interactions
has to return to the small scales in order 
to be dissipated by viscosity. In fact, there is
no energy sink on the large scales that could take the energy out of the system.
Therefore  the condition for establishing a 
stationary state is that 
the time average energy transfer  
has to be zero on all wavenumbers, 
and dissipation has to equal bubble forcing,
i.e., $T(k)=0$ and $D(k) = F_b(k)$, where the time dependence has  
dropped out after the average on time.

As we show in Fig.~\ref{spe_forc_eps}$a$, apart from the large scales where
the average still has not converged, this requirement is satisfied by
our simulation.
In real flow, with walls instead of periodic boundary conditions,
energy dissipation in the developing boundary layers will of course
eventually play a crucial role in the energy balance. 

The time evolution of the energy spectrum can be 
compared to the one presented by \cite{mur00c}, 
where the authors study a similar system, namely fluid motion 
generated by rising bubbles, by applying a different 
technique for the implementation of two-way coupling. 
The results agree qualitatively, i.e., the initial induction 
of structures at large scale is followed by 
a state in which the slope of the energy spectrum is reduced. 
As the authors state themselves, the reason is connected with the
temporal evolution of the bubble distribution.
Indeed, bubble clusters, that are assembled in the beginning
and are able to force the liquid efficiently, are
not stable and bubbles tend to distribute uniformly in the flow.
Moreover, the structures induced in the flow itself
are far too weak to trap the bubbles. Thus, 
the phenomenon of vortex trapping of bubbles does not occur and
therefore high local bubble concentration 
as in bubbly turbulent flows (see e.g.\ \cite{maz03a,cal08,cal08b})
are not created here.

Note that the front-tracking simulations of ref.\ \cite{bun02b} 
for 27 bubbles with a bubble Reynolds number of about 25 gave a quite
different spectrum, namely a slope of about $-3.6$ in the large wavenumber
regime. Indeed, a slope $-3$ has theoretically 
been attributed \cite{lan91} to this regime, 
in which the energy deposited by the wake is directly dissipated and which
obviously cannot be modelled with the point-particle approach.
All this wake energy dissipation is missing in our approach.

\begin{figure}
\begin{center}
\includegraphics*[width=5cm]{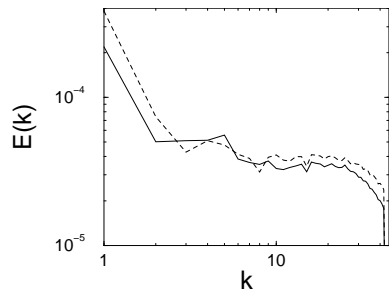}
\caption{Energy spectra, in the statistically stationary state, 
for case $(a)$ (dashed line) and $(d)$ (solid line) of Table \ref{table1}.}
\label{spe_final_eps}
\end{center}
\end{figure}

We carry on the comparison with the results of 
\cite{mur00c}, by looking at the
high wavenumbers behavior of the spectrum.
In that paper a steeper slope with respect to 
homogeneous and isotropic 
turbulence is observed. The critical wavenumber $k_c$ above which 
it shows up is estimated by the average
distance between the bubbles, that is 
$L_c \sim 2\pi/N_b^{1/3}$.
In our simulation, for case $(a)$ of Table \ref{table1}, we have:
$L_c \sim 0.12$ 
(about $1/50$ of the box with $L_0 = 2 \pi$), 
thus $k_c = 2\pi / L_c \sim 52$ and for case 
$(d)$: $L_c \sim 0.17$, thus $k_c \sim 37$. 
As we show in Fig. \ref{spe_final_eps}, 
a transition
in the slope of the energy spectrum occurs 
at high wavenumbers. However, neither the 
critical wavenumber nor the slope of the spectra can be
clearly defined.

We find agreement with the results 
of \cite{mur00c} on the strongly anisotropic energy
distribution along the three velocity components. Indeed, also in that
work, about the $90\%$ of the flow energy is contained in
the vertical component $(z)$ of the fluid velocity.

We again also compare our results with the case for sedimenting particles in originally still fluid \cite{faeth,mizu}:
Also for that case the energy spectrum is very broad. The spectral slope of the frequency spectrum
was found to be  around $-1.1$ for remarkably several decades, but a comparison of this value with the slopes in the
wavenumber spectra 
obtained in this paper is difficult as Taylor's frozen flow hypothesis will most likely not hold for this weak turbulence.

\section{Physical explanation of the results}\label{sec5}
The occurrence of the inverse cascade phenomenon in three dimensional 
turbulence has been related to the presence of 
strong anisotropies at small scales (see e.g.\ \cite{yak87,hef89}). 
These anisotropies can be produced by bubble clusters 
elongated in the gravity direction. 
Within them, the energy production term
due to the bubbles 
can be far more intense in the vertical direction than in 
the horizontal ones, owing to the high values reached by the
$\langle {\bf u} \cdot {\bf g} \rangle$ contribution.
The stability of these structures is opposed by 
horizontal forces that laterally spread the bubbles.
When considering the bubble motion equation, it appears that the lift
is the most relevant of such forces.
Therefore we further investigate the system
by comparing the outcome of simulations
with lift and without lift force.

Surprisingly, 
without lift 
the results are very  different. 
In Fig.~\ref{fig1}$b$ the total energy 
in the two simulations, one including the lift (solid line) and
the other excluding it (dashed line), are compared. In both cases
the bubbles parameters correspond to run $(a)$ in Table \ref{table1}.
The energy induced in the second simulation is up to $30$
times larger than in the first.
Also Murai {\it et al.} \cite{mur00a} measured a higher
turbulence intensity in numerical simulations without lift force
than in simulations with lift,
though the difference detected is quantitatively much smaller than
in our case.

The behavior in spectral space is remarkably different, too.
In Fig.~\ref{fig3}$b$ the time evolution of 
the energy spectrum in the latter simulation is presented. 
The spectrum is averaged on four subsequent time 
intervals, which are the same as for Fig \ref{fig3}$a$.
The process of inverse energy cascade is now strongly enhanced.
In fact, the spectral intensity at small wavenumbers 
increases in time, whereas it is
constant at large ones.

Moreover, it is remarkable that a local slope close to $-5/3$ settles nearly at 
once at high wavenumbers and is stable during the whole process -- though the scaling regime again is not very pronounced
due to the small Taylor-Reynolds numbers. 
Therefore the small scale forcing
is strong enough  
to generate a flow that presents the same characteristics as 
real three-dimensional turbulence -- or also two-dimensional turbulence
in the inverse cascade regime. 
On the other hand, this simulation is not statistically stationary. 
In Fig.~\ref{spe_forc_eps}$b$, we show that there is a
difference on a wide range of scales between
the bubbles forcing term $F_b(k)$ and the fluid viscous
dissipation $D(k)$. Thus the condition of stationarity is
not fulfilled and the large flow scales are still fed 
with energy from the small ones. 

We again stress that the model for the equation of 
motion without lift force does not give a complete representation
of the surface forces acting on bubbles or particles of $Re \sim O(1)$.
Indeed, previous works (see e.g.\  refs.\
\cite{leg98,maz03b} and ref.\ \cite{dan90}  for the case of uniform shear) have pointed out the relevance of
the lift in this regime. 
Of course our results can only expected to be qualitative and not 
quantitative,
as the near-field interactions between the bubbles is not correctly
described by the present point-particle 
 approach and as the lift coefficient is set to be constant,
$C_L= 1/2$. 
However, refined expressions for the lift force are not likely to
give qualitatively different behaviors.
The main effect of the lift is to cause the
bubble dispersion along the horizontal directions, thus strongly
reducing the anisotropy in the flow caused by the
forcing in the vertical direction.
Indeed, by definition, the lift force drifts 
the bubbles in horizontal planes,
in directions perpendicular to their average motion. \\

\begin{figure}
\begin{center}
\includegraphics*[width=5.5cm]{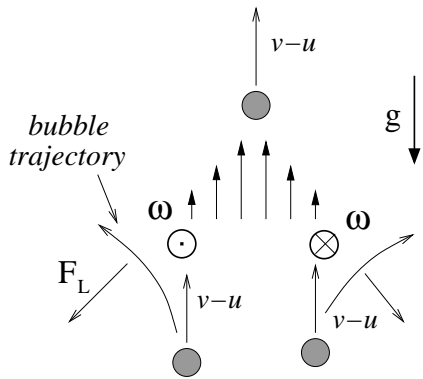}
\caption{Sketch of the action of the lift force on a bubble rising in the wake
of another one: the lift tends to expel the bubble from the wake.}
\label{koch_eps}
\end{center}
\end{figure}

We now qualitatively investigate the breaking effect of the lift on
vertical bubble chains. Note again that for a quantitative analysis
the near-field as e.g.\ obtained from the front-tracking simulations of refs.\
\cite{bun02a,bun02b,esm99,esm05} is crucial. 
We base our analysis on the description of the two-bubble 
long range interactions proposed in ref.\ \cite{koc93}.
The main finding is that a bubble in the wake of another one 
experiences, because of the lift, a lateral force, leading to 
a deficit of nearby bubbles in the gravity direction 
(see Fig.\ \ref{koch_eps}).

\begin{figure} 
\begin{center}
\includegraphics*[width=6cm]{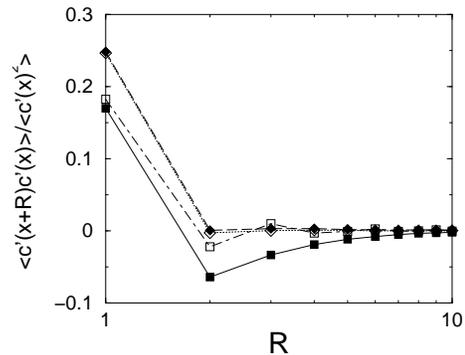}
\caption{Autocorrelation function $\rho_{12}({\bf R})$ as a function of the 
distance $|{\bf R}|$, in the horizontal $x$-$y$ directions (open symbols)
and in the vertical $z$ direction (filled symbols). The results indicate
simulations without lift force (diamonds) and with lift force (squares).}
\label{pairdistr_eps}
\end{center}
\end{figure}

We further explore this phenomenon
by computing the bubble density autocorrelation function, defined
according to:
\begin{equation}
\rho_{12}({\bf R}) = { \langle c'({\bf x} + {\bf R} )c'({\bf x}) \rangle 
\over \langle c'({\bf x})c'({\bf x}) \rangle}
\label{pairdistr}
\end{equation} 
Here $c'({\bf x})$ is the fluctuation of the bubble concentration
in ${\bf x}$ with respect to the average value $\alpha$, and the brackets
denote averages over all ${\bf x}$. 
We consider the autocorrelation in the horizontal, $(x$-$y)$, plane
and in the vertical, $(z)$, direction separately.
The analysis is carried out for the two simulations presented
in Fig.\ \ref{fig1}$b$.
The results are plotted in Fig.\ \ref{pairdistr_eps}. It is shown that,
whereas in the simulation without lift forces the pair correlation goes 
monotonically to zero, in the one with lift 
the autocorrelation in the $z$-direction becomes negative 
and later on approaches zero from below. A negative
autocorrelation at small distances $R$ is detected
also in the horizontal planes. The interpretation of 
the result is that the bubbles approach is resisted by the 
lift, and this is occurring especially in the vertical direction, 
where a bubble rising in the wake of another experiences 
horizontal forces that expels it from the wake.

Qualitatively, the organization of the bubbles in our simplified 
simulations
 is similar to what has been observed in above
mentioned front tracking simulations by Tryggvason and coworkers
\cite{esm98,esm99,bun02a,bun02b,esm05}
or in the experiments of ref.\
\cite{car01}: Small bubbles with $Re \sim 1$ were dispersed in a 
nearly homogeneous manner, with an increasing tendency of horizontal alignment
when the bubble Reynolds number 
approaches 10. 
Also in the recent experiments 
by Harteveld {\it et al.}~\cite{har03b} 
bubbly driven flows have been found to be rather homogeneous 
and no vortex trapping has been detected.
The front-tracking simulations show that 
for even larger bubble Reynolds numbers horizontal bubble clusters 
can emerge, which also have been seen in the experiments of \cite{zen01}. 
Even in the two-dimensional simulations described in 
ref.~\cite{esm96} a preferential 
tendency of bubbles
to stay rather side-by-side (along the horizontal direction)
than in ``tandem configuration'' (along the vertical) was reported.

\section{Conclusions}\label{sec6}
The behavior of a flow driven exclusively by rising bubbles  
has been investigated by direct numerical simulation for the Navier 
Stokes equations and Lagrangian tracking for the 
bubble trajectories, where the bubbles have been treated as point particles
on which effective forces act.
The evolution of global quantities, like the total flow
energy, as well as of spectral quantities has been followed in time.
The results show that the bubble motion initially generates large scale
structures by local in scale energy transfer, though the corresponding $-5/3$ scaling regime in the spectrum is not very pronounced 
due to the small Taylor-Reynolds numbers. 
Later on, however, the bubbles distribution 
tends to be more disperse, the energy spectrum becomes flat
and energy input equal viscous dissipation at all scales. 
Therefore, the statistically stationary state of this
pseudo-turbulent velocity field does not possess the
characteristics of real turbulence. 

We give qualitative 
evidence that the physics that determines it are the 
bubble-bubble indirect interactions that occur via the carrier flow.
Indeed, a bubble in the wake of another one, experiences, because of
the lift, a horizontal force that prevents the assembling and
the stability of vertical clusters, similarily as has been
observed in the front-tracking simulations of bubbles of comparable
size \cite{esm98,esm99} or of larger bubbles
 \cite{bun02a,bun02b}, where the bubbles also distribute homogeneously
and horizontal pairs of bubbles are favored. 
In our case,
as a consequence the total forcing induced in the flow
is not strong enough to sustain high energy levels and an inverse
energy cascade from small to large scales.

The results presented in this work 
apply to a flow with periodic boundary conditions. In a real experiment 
such as in Taylor-Couette flow,
the existence of boundaries  can lead to the generation
of large scale vortex structures, that, in turn, affect the bubble
motion, as seen in the experiments of Murai {\it et al.}~\cite{mur05} and
the corresponding numerical simulations of Sugiyama {\it et al.}~\cite{sug08b}.
In those flows the energy dissipation in the developing boundary layers
will play a crucial role in the stationary energy balance.

We understand this work as complementary to the front-tracking simulations:
Bunner and Tryggvason \cite{bun02b} end their numerical study on the
dynamics of homogeneous bubble flows with the question: ``What happens
when the number of bubbles increases beyond 216?'' (the number of bubbles
they could numerically treat in the paper). Here we treat up to 288000 bubbles,
though in an approximate and simplified way. Nonetheless, we see very similar
phenomena as observed in the front-tracking simulations, and can even
identify the lift force as origin of the homogeneous bubble distribution,
by artifically turning it off. A full verification can of course only come
from simulations which {\it both} resolve the individual bubbles including their
wakes {\it and} deal with hundred-thousands of individual bubbles. Such 
simulations will however unfortunately not be numerically 
feasible for many years to come.

{\it Acknowledgements:}
The authors thank A.\ Prosperetti and L.\ van Wijngaarden
for discussions and comments on the manuscript.


\end{document}